\documentclass[vecphys]{svmult}
\usepackage{makeidx}
\usepackage{graphicx}
\usepackage{multicol}

\makeindex

\begin{document}

\title*{NMR and $\mu$SR in Highly Frustrated Magnets}
% Use \titlerunning{Short Title} for an abbreviated version of
% your contribution title if the original one is too long
\author{Pietro Carretta\inst{1} and Amit Keren \inst{2}}
% Use \authorrunning{Short Title} for an abbreviated version of
% your contribution title if the original one is too long
\institute{1- Dipartimento di Fisica "A.Volta" - University of Pavia - Via Bassi, 6 - 27100 Pavia (Italy)
\texttt{carretta@fisicavolta.unipv.it}\\ 2- Technion-Israel Institute of Technology, Physics Department, Haifa
32000, Israel \texttt{keren@physics.technion.ac.il} }
%
% Use the package "url.sty" to avoid
% problems with special characters
% used in your e-mail or web address
%
\maketitle

Hereafter we shall present a brief overview on some of the most significant achievements obtained by means of NMR
and $\mu$SR techniques in highly frustrated magnets. First the basic quantities measured by the two techniques
will be presented and their connection to the microscopic static and dynamical spin susceptibility recalled. Then
the main findings will be outlined, starting from the most simple frustrated units, the molecular nanomagnets, to
artificially built frustrated systems as $^{3}$He on graphite, to magnets with a macroscopically degenerate
ground-state as the ones on a pyrochlore or kagom\'{e} lattices.

\section{Some basic aspects of NMR and $\protect\mu$SR techniques}

\label{sec:1}

NMR and $\mu $SR are very powerful techniques which allow to investigate the microscopic properties of spin
systems through the study of the time evolution of the nuclear magnetization $\vec{M}(t)$ and of the muon spin
polarization $\vec{P}(t)$, respectively \cite{Slichter,Schenck}. Each technique has its advantages and
disadvantages. In NMR one knows the crystallographic position of the nuclei under investigation and, therefore NMR
results can be more suitably compared to theories. On the other hand, NMR experiments cannot be performed in
compounds where just low sensitivity nuclei are present or where the fast nuclear relaxations prevent the
observation of an NMR signal. Still, polarized muons can be injected into the sample and used as a probe of the
local microscopic properties of the system under investigation. Moreover, by means of $\mu$SR it is possible to
detect relaxation times shorter than 0.1 $\mu $s, about two order of magnitudes shorter than the shortest
relaxation time NMR can measure. Since the nuclear magnetization is the quantity detected in the NMR experiments,
generally a magnetic field has to be applied to generate it. On the other hand, the muon beam is already polarized
before entering the sample, so that the system under investigation can also be studied in zero field by means of
$\mu$SR. This aspect is particularly relevant if one wants to investigate the intrinsic properties of a certain
system without perturbing it with a magnetic field. Nevertheless, novel ground-states can be induced by the
application of high magnetic fields (typically above 10 Tesla) where $\mu$SR experiments cannot be performed while
NMR experiment can. Hence, although both techniques appear to measure similar quantities (see next section), in
view of the above considerations they are often complementary and their combination is a rather powerful method to
investigate the local microscopic properties of frustrated magnets.

\subsection{Line shift and line width}

\begin{figure}[h]
\vspace{6cm} \includegraphics{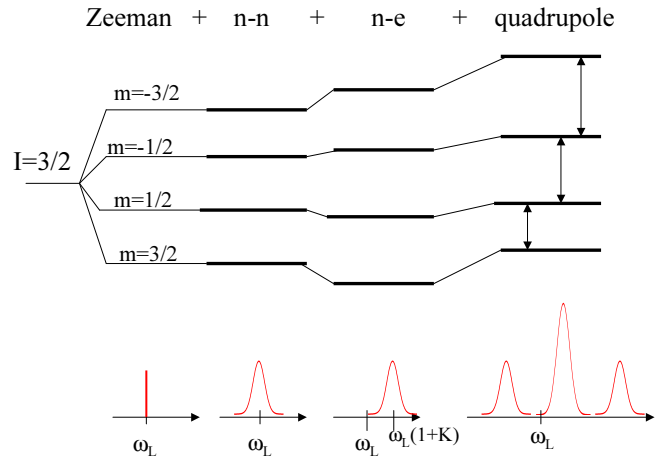} \caption{Schematic
illustration of the modifications in the hyperfine levels of $I=3/2$ nuclei, due to the different terms of the
nuclear hyperfine Hamiltonian. The corresponding modifications in the NMR spectra are reported at the bottom of
the figure. } \label{fig:10}
\end{figure}

The time evolution of $\vec{M}(t)$ and $\vec{P}(t)$ is determined by the hyperfine interactions which can be
summarized in the form
\begin{equation}
\mathcal{H}= \mathcal{H}_{z}+ \mathcal{H}_{n-n}+ \mathcal{H}_{n-e}+ \mathcal{H}_{EFG}\,\,\,.  \label{GlobalHamil}
\end{equation}
The effect of all the four terms on the hyperfine levels and on the NMR spectra for $I=3/2$ are depicted in Fig.
\ref{fig:10}. We recall that the intensity of the whole spectrum is proportional to the nuclear magnetization.
%%%%%%%%%%%%%%%%%%%%FIG2%%%%%%%%%%%%%%%%%%%%%%%%%%%%%%%%%%%%%%%%%%%%
%%%%%%%%%%%%%%%%%%%%%%%%%%%%%%%%%%%%%%%%%%%%%%%%%%%%%%%%%%%%%%%%%%%%
The first term describes the Zeeman interaction with an external field,
while $\mathcal{H}_{n-n}$ is the dipole-dipole interaction among the nuclear
spins or between the muon and the nuclear spins. This interaction yields a
broadening of the NMR \cite{Slichter} and $\mu $SR \cite{Schenck} spectra.
In certain compounds, as the cuprates, the nuclear dipole-dipole interaction
is mediated by the electron spins and from the dipolar broadening
information on the static electron spin susceptibility $\chi ^{\prime }(\vec{%
q})$ is obtained \cite{Slic2}. The last term $\mathcal{H}_{EFG}$ describes the interaction between the nuclear
electric quadrupole moment $Q$ and the electric field gradient (EFG) generated by the charge distribution around
the nucleus. This term is non-zero when the nuclear spin $I>1/2$ and is, of course, absent in the muon interaction
Hamiltonian. The quadrupole interaction is very sensitive to the modifications in the local configuration and
allows to evidence distortions induced by the spin-lattice coupling \cite{OferPRB06} or the presence of a
non-homogeneous charge distribution induced by charge ordering, for instance \cite{Alloul}. The third term is the
most relevant one to probe frustrated magnetism, as it describes the hyperfine interaction with the electron spins
$\vec{S}$. As most of the systems we shall be dealing with in the following are insulators, one can write
\begin{equation}
\mathcal{H}_{n-e}=-\gamma \hbar \sum_{i,k}\vec{I}_{i}\tilde{A}_{ik}\vec{S}%
_{k}
\end{equation}%
with $\tilde{A}_{ik}$ the hyperfine coupling tensor, $I$ and $S$ are the
nuclear/$\mu ^{+}$ and electronic spin operator respectively, $i$ and $k$
are the nuclear/$\mu ^{+}$ and electron spin indexes, respectively, while $%
\gamma $ the nuclear ($\mu ^{+}$) gyromagnetic ratio. Then, the hyperfine
field at the $i$-th nucleus/$\mu ^{+}$ will be given by $\vec{h}_{i}=\sum_{k}%
\tilde{A}_{ik}\vec{S}_{k}$ and in the presence of a non-zero average
polarization $<\vec{S}>$\thinspace \thinspace\ $\vec{h}_{i}=\sum_{k}\tilde{A}%
_{ik}<\vec{S}_{k}>$. Thus, one can directly estimate $<\vec{S}>$ from the
precessional frequency $\omega =\gamma \sum_{k}\tilde{A}_{k}<\vec{S}_{k}>$
of the nuclei or of the muons around the local field.

\begin{figure}[h]
\vspace{5cm} \includegraphics{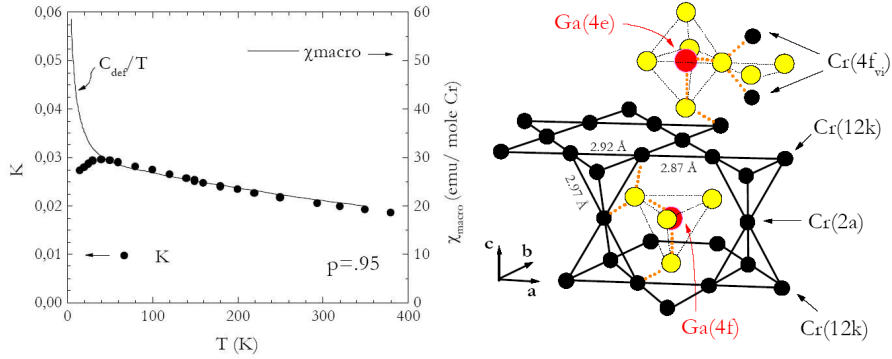} \caption{Temperature
dependence of $^{71}$Ga(4f) NMR shift $K$ in SrCr$_{9p}$Ga$_{12-9p}$O$_{19}$ ($p=0.95$) compared to the
macroscopic susceptibility derived with a magnetometer. Since $^{71}$Ga(4f) nuclei are strongly coupled only to
chromium ions within the kagom\'e bilayer one can single out just the intrinsic susceptiblity of the latter,
whereas the macroscopic susceptibility detects also the contribution from defects.\cite{Limot}} \label{fig:1L}
\end{figure}

When an external field $\vec{H}_{0}\parallel \hat{z}$ is applied the local magnetic field becomes
\begin{equation}
\vec{B}=\vec{H}_{0}+\sum_{k}\tilde{A}_{k}<\vec{S}_{k}>
\end{equation}%
and the resonance frequency will be shifted to
\begin{equation}
\omega =\omega _{L}(1+{K})
\end{equation}%
with $\omega _{L}=\gamma H_{0}$ the Larmor frequency and, for $H_{0}\gg
|\sum_{k}\tilde{A}_{k}<\vec{S}_{k}>|$,
\begin{equation}
{K}=\frac{\biggl(\sum_{k}\tilde{A}_{k}<\vec{S}_{k}>\biggr)_{z}}{H_{0}}\,\,.
\end{equation}%
In general $K$ is a tensor and ${\tilde{K}}=\sum_{k}\tilde{A}_{k}\tilde{\chi}%
(\vec{q}=0,\omega =0)$. Hence, one notices that from the shift in the precessional frequency of the nuclei (or
$\mu ^{+}$) one can derive the static uniform susceptibility associated only with those electron spins which are
coupled to the nuclei under investigation see Fig.\ref{fig:1L}.
%%%%%%%%%%%%%%%%%%%%FIG2%%%%%%%%%%%%%%%%%%%%%%%%%%%%%%%%%%%%%%%%%%%%
%%%%%%%%%%%%%%%%%%%%%%%%%%%%%%%%%%%%%%%%%%%%%%%%%%%%%%%%%%%%%%%%%%%%

If, for some reason, the distance between nuclei or muon and the electronic spins is random or there are missing
spins in the sample, the hyperfine coupling $\tilde{A}$ will be a random variable leading to a distribution of
precessional frequencies and to an increase in the relaxation of the polarization/magnetization. For an external
field applied perpendicular to the initial polarization of the muons one finds
\begin{equation}
P_{\bot }(t)= \exp \left( -\left[ \frac{t}{T_{2}^{\ast }}\right] ^{2}\right) \cos (\omega t).
\label{TFRelaxation}
\end{equation}%
$1/T_{2}^{\ast }$ could also represent the decay rate of the NMR signal after an RF pulse. Assuming a distribution
of hyperfine fields in the $\hat{z}$ direction one can write $\tilde{A}_{k}$ as a sum of a mean value
$\overline{A_{k}}$ plus a fluctuating component $\delta A_{k}$. For the distribution
\[
\rho (\delta A_{k})=\frac{1}{\sqrt{2\pi }\sigma _{k}}\exp \left( -\frac{%
\delta A_{k}^{2}}{2\sigma _{k}^{2}}\right) ,
\]%
one finds that
\begin{equation}
\frac{1}{{T_{2}^{\ast }}}=\gamma \left( \sum_{k}\sigma _{k}^{2}\right) ^{1/2}\chi H_{0}  \label{T2toChi}
\end{equation}%
is the width of the spectrum which has an average shift
\begin{equation}
K=\frac{\omega -\gamma H_{0}}{\gamma H_{0}}=\chi \sum_{k}%
\overline{A_{k}}.  \label{KtoAandChi}
\end{equation}%
If $\sigma _{k}$ and $\overline{A}$ are temperature independent parameters
we expect
\begin{equation}
1/T_{2}^{\ast }\propto K  \label{T2toK}
\end{equation}%
where the temperature is an implicit parameter. It is noticed that this
proportionality is correct regardless of the form of the relaxation (\ref%
{TFRelaxation}). The breakdown of the validity of Eq.~\ref{T2toK} would indicate a modification in the hyperfine
couplings, which is typically expected when lattice distortions take place (see Sec's. 2.2 and 2.5). On the other
hand, in certain systems, although the hyperfine coupling is constant, the spin polarization can be
site-dependent. Then, Eq.\ref{T2toK} no longer holds and the line broadening and its shape reflects the
distribution of the local spin polarization.\cite{Tedoldi}

\subsection{Nuclear and muon spin lattice relaxation rate $1/T_{1}$}

The transitions among the hyperfine levels, driven by the time dependent part of the hyperfine Hamiltonian (not
shown in Eq.~\ref{GlobalHamil}), modify the nuclear spin population on each level, namely the longitudinal
component of nuclear magnetization. Thus, in a frustrated magnet from the time evolution of the nuclear
magnetization it is possible to derive information on the spin dynamics, which drives the fluctuations of the
hyperfine field. The recovery of the longitudinal component of the nuclear magnetization, after the nuclear spin
ensemble has been brought out of equilibrium with an $ad$ $hoc$ RF pulse sequence, is described by a
characteristic decay rate $1/T_{1}$. In case of relaxation mechanisms driven by fluctuations of the hyperfine
field $\vec{h}(t)$, by resorting to
time-dependent perturbation theory ($h$ is considered small with respect to $%
H_{0}$) and assuming that the frequency of the field fluctuations $\omega= 2\pi\nu \gg 1/T_{1}$, one can write
\begin{equation}
\frac{1}{T_{1}}=\frac{\gamma ^{2}}{2}\int_{-\infty }^{+\infty }e^{i\omega
_{L}t}<h_{+}(t)h_{-}(0)>dt\,\,\,.  \label{BasicT1}
\end{equation}%
This fundamental expression shows that $1/T_{1}$ is driven by the transverse components of the fluctuating field
at the nucleus, owing to magnetic-dipole selection rules, and that $1/T_{1}$ is proportional to the Fourier
transform of the correlation function at the resonance frequency, in order to allow for energy conservation. In
other terms $1/T_{1}$ probes the spectral density at the resonance frequency $\omega_{L}$ which is typically in
the MHz range, orders of magnitude below the spectral range accessed by inelastic neutron scattering experiments.
It should be noticed that this does not mean that from $1/T_1$ one cannot estimate relevant energy scales much
larger than $\hbar\omega_L$. In fact, when sum rules apply, as it is often the case in spin systems, the amplitude
of the low-frequency spectral density is determined by the characteristic frequency of the fluctuations $\omega>>
\omega_L$.

%%%%%%%%%%%%%%%%%%%%FIG2%%%%%%%%%%%%%%%%%%%%%%%%%%%%%%%%%%%%%%%%%%%%
\begin{figure}[h!]
\vspace{6cm} \includegraphics{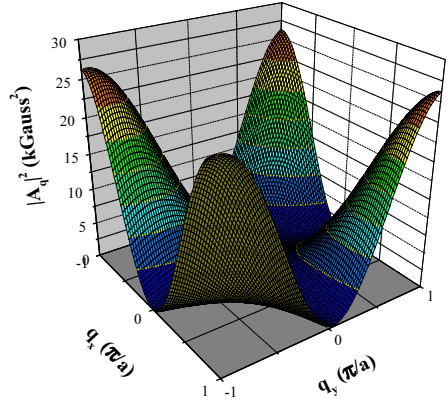}
\caption{$^{29}$Si form factor in the first Brillouin zone of the
two-dimensional frustrated antiferromagnet Li$_2$VOSiO$_4$. It is evident
that excitations at wave-vectors $(\pm \protect\pi/a,0)$ or $(0, \pm \protect%
\pi/a)$ are filtered out, namely that $^{29}$Si $1/T_1$ is not sensitive to
these modes.}
\label{Siform}
\end{figure}
%%%%%%%%%%%%%%%%%%%%%%%%%%%%%%%%%%%%%%%%%%%%%%%%%%%%%%%%%%%%%%%%%%%%%%%%%%%%%%

In general, when collective spin excitations are present one can write
\begin{equation}
\vec{h}(t)=\frac{1}{\sqrt{N}}\sum_{\vec{q}}\sum_{k}e^{i\vec{q}\vec{r}_{k}}%
\tilde{A}_{k}\vec{S}(\vec{q},t)
\end{equation}%
and by substituting the transverse components of $\vec{h}(t)$ in Eq. 10 expression one finds that
\begin{equation}
\frac{1}{T_{1}}=\frac{\gamma ^{2}}{2}\frac{1}{N}\sum_{\vec{q},\alpha=x,y,z}\biggl(|A_{%
\vec{q}}|^{2}S_{\alpha \alpha }(\vec{q},\omega _{L})\biggr)_{\perp }\,\,\,.
\label{T1S}
\end{equation}%
One notices that, being the nuclei local probes, $1/T_{1}$ is related to the integral over the Brillouin zone of
the component of the dynamical structure factor $S_{\alpha \alpha }(\vec{q},\omega _{L})$ at the Larmor frequency.
In Eq. 12 $|A_{\vec{q}}|^{2}$ is the form factor which gives the hyperfine coupling of the nuclei with the spin
excitations at wave-vector $\vec{q}$ (see Fig. \ref{Siform}). The term $\perp $ indicates that one has to consider
the products $|A_{\vec{q}}|^{2}S_{\alpha \alpha }(\vec{q},\omega _{0})$ associated with the perpendicular
components of the hyperfine field at the nucleus. From the fluctuation-dissipation theorem, by recalling that
usually $k_{B}T\gg \hbar \omega _{L}$ one can also write
\begin{equation}
\frac{1}{T_{1}}=\frac{\gamma ^{2}}{2}\frac{k_{B}T}{\hbar }\frac{1}{N}\sum_{\vec{q},
\alpha=x,y,z}\biggl(|A_{\vec{q}}|^{2}\frac{\chi "_{\alpha \alpha }(\vec{q},\omega _{L})}{\omega
_{L}}\biggr)_{\perp } \label{T1NMR}
\end{equation}%
This rather general expression shows how the nuclear or muon spin-lattice
relaxation is related to the spectrum of the excitations characteristic of
each frustrated magnet.

The above equations apply also to $\mu$SR spin-lattice relaxation rate when a large magnetic field is applied. On
the other hand, when $\mu $SR operates in zero field, standard perturbation methods to analyze the data, such as
Eq. \ref{BasicT1}, are no longer valid. This is because the transverse direction is not defined and the internal
field is not small compared to $H$. Accordingly, different methods are required to account for the muon relaxation
function in zero and small external fields. Moreover, usually one does not know the muon stopping site, and
therefore the discussion is done using the field at the muon site $B$ rather than the hyperfine coupling $A$. The
treatment of the muon $T_{1}$ in zero or small
fields is done in two steps. The first step is the static case where $%
T_{1}=\infty $. In the second step, the dynamic fluctuations are added and $%
T_{1}$ becomes finite.

\subsection{$\mu$SR: the static case}

\begin{figure}[h!]
\vspace{3.5cm} \includegraphics{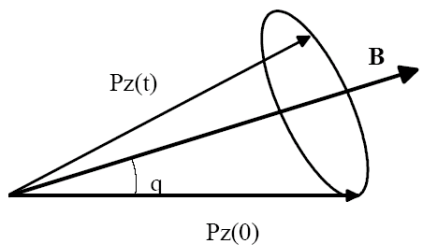}
\caption{Muon spin polarization rotating around a magnetic field in an
arbitrary direction.}
\label{Static1}
\end{figure}

The fully polarized muon, after entering the sample, comes to rest in a magnetic environment. Since the mechanism
which stops the muon is much stronger than any magnetic interaction, the muon maintains its polarization while
losing its kinetic energy. Once the muon reaches its site, the muon spin starts to evolve in the local field
$\mathbf{B}$. The muon polarization $P_{z}$ along the $\mathbf{\hat{z}}$ direction is given by the double
projection expression
\begin{equation}
P_{z}(\mathbf{B},t)=\cos ^{2}\theta +\sin ^{2}\theta \cos (\gamma _{\mu
}\left\vert \mathbf{B}\right\vert t)  \label{Gz}
\end{equation}%
where $\theta $ is the angle between the initial muon spin and the local
field direction (see Fig.~\ref{Static1}). This angle is related to the field
values by
\[
\cos ^{2}\theta =\frac{B_{z}^{2}}{\mathbf{B}^{2}}\,\,\,\,\,,\,\,\,\,\,\sin
^{2}\theta =\frac{B_{x}^{2}+B_{y}^{2}}{\mathbf{B}^{2}}\,\,\,.
\]

%%%%%%%%%%%%%%%%%%%%FIG2%%%%%%%%%%%%%%%%%%%%%%%%%%%%%%%%%%%%%%%%%%%%
%%%%%%%%%%%%%%%%%%%%%%%%%%%%%%%%%%%%%%%%%%%%%%%%%%%%%%%%%%%%%%%%%%%%%%%%%%%%%%

In a real sample, however, there will be a distribution of internal fields
and the averaged polarization is
\begin{equation}
\overline{P}_{z}(t)=\int \rho (\mathbf{B})\left[ \frac{B_{z}^{2}}{\mathbf{B}%
^{2}}+\frac{B_{x}^{2}+B_{y}^{2}}{\mathbf{B}^{2}}\cos (\gamma _{\mu
}\left\vert \mathbf{B}\right\vert t)\right] d^{3}\mathbf{B}  \label{StatInt}
\end{equation}%
where $\overline{P}_{z}(t)$ is the sample averaged polarization, and $\rho (%
\mathbf{B})$ is the field distribution which is normalized according to $%
\int \rho (\mathbf{B})d\mathbf{B}^{3}=1$. If the distribution of internal
fields is only a function of $\left\vert \mathbf{B}\right\vert $ then we can
write
\[
\overline{P}_{z}(t)=\int \rho (\left\vert \mathbf{B}\right\vert )\left[ \cos
^{2}\theta +\sin ^{2}\theta \cos (\gamma _{\mu }\left\vert \mathbf{B}%
\right\vert t)\right] B^{2}dBd\Omega .
\]%
It is convenient to define $\rho ^{\prime }(\left\vert \mathbf{B}\right\vert
)=4\pi \rho (\left\vert \mathbf{B}\right\vert )$, so that $\int \rho
^{\prime }(\left\vert \mathbf{B}\right\vert )\mathbf{B}^{2}dB=1$ and the
angular dependence can be integrated out giving%
\[
\overline{P}_{z}(t)=\frac{1}{3}+\frac{2}{3}\int \rho ^{\prime }(\left\vert
\mathbf{B}\right\vert )\cos (\gamma _{\mu }\left\vert \mathbf{B}\right\vert
t)B^{2}dB.
\]%
If, for example, the system has long range order the field at the muon site is centered around a specific value
$\omega _{0}/\gamma _{\mu }$ so that
\[
\rho ^{\prime }(\left\vert \mathbf{B}\right\vert )=\frac{\gamma _{\mu }}{%
\sqrt{2\pi }\Delta \mathbf{B}^{2}}\exp \left[ -\gamma _{\mu }^{2}\left(
\left\vert \mathbf{B}\right\vert -\frac{\omega _{0}}{\gamma _{\mu }^{2}}%
\right) ^{2}/2\Delta ^{2}\right] ,
\]%
then
\[
\overline{P}_{z}(\omega _{0},\Delta ,t)=\frac{1}{3}+\frac{2}{3}\exp \left( -%
\frac{\Delta ^{2}t^{2}}{2}\right) \cos (\omega _{0}t)
\]%
and oscillations will be observed in the data. At long time the polarization
will relax to $1/3$ since effectively $1/3$ of the muons experience a field
parallel to their initial spin direction and do not relax.

%%%%%%%%%%%%%%%%%%%%FIG2%%%%%%%%%%%%%%%%%%%%%%%%%%%%%%%%%%%%%%%%%%%%
\begin{figure}[h!]
\vspace{2.5cm} \includegraphics{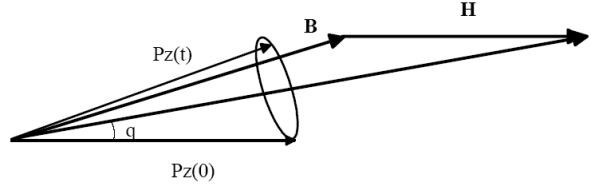}
\caption{Muon spin polarization rotating around the vector sum of an
external magnetic field in the initial muon polarization direction, and an
arbitrary internal field.}
\label{Static2}
\end{figure}
%%%%%%%%%%%%%%%%%%%%%%%%%%%%%%%%%%%%%%%%%%%%%%%%%%%%%%%%%%%%%%%%%%%%%%%%%%%%%%

When a longitudinal field $H$ is applied in the direction of the initial
muon spin as in Fig.~\ref{Static2}, the situation becomes more complicated,
and there is no closed form expression. However, some simplifications could
be made to reduce the dimension of the integrals for the purpose of
numerical calculations. For example, if the local field is completely random
with a Gaussian distribution then
\begin{equation}
\rho (\mathbf{B})=\frac{\gamma _{\mu }^{3}}{(2\pi )^{3/2}\Delta ^{3}}\exp
\left( -\frac{\gamma _{\mu }^{2}[\mathbf{B-}H_{0}\mathbf{\hat{z}}]^{2}}{%
2\Delta ^{2}}\right) .  \label{Rho}
\end{equation}

In this case Eq.~\ref{StatInt} could be simplified to \cite{HayanoPRB79}
\begin{eqnarray}
\overline{P}_{z}(\omega_{\mathrm{L}},\Delta,t)=1-\frac{2\Delta^{2}}{(\omega_{%
\mathrm{L}})^{2}}\left[ 1-\exp(-\frac{1}{2}\Delta^{2}t^{2})\cos(\omega_{%
\mathrm{L}}t)\right] +  \nonumber \\
\frac{2\Delta^{4}}{(\omega_{\mathrm{L}})^{3}}\int_{0}^{t}\exp(-\frac{1}{2}%
\Delta^{2}\tau^{2})\sin(\omega_{\mathrm{L}}\tau)d\tau  \label{GLF}
\end{eqnarray}
%%%%%%%%%%%%%%%%%%%%FIG2%%%%%%%%%%%%%%%%%%%%%%%%%%%%%%%%%%%%%%%%%%%%
\begin{figure}[h!]
\vspace{6cm} \includegraphics{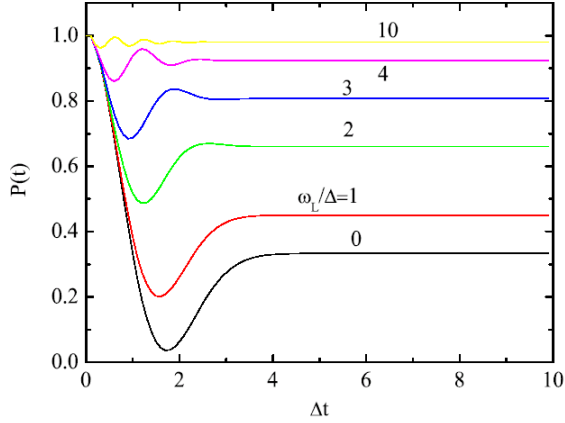}
\caption{Muon polarization function in a Gaussian internal field
distribution and external field pointing in the initial muon spin direction.
Different values of the external field $H$ are shown.}
\label{SGLFKT}
\end{figure}
%%%%%%%%%%%%%%%%%%%%%%%%%%%%%%%%%%%%%%%%%%%%%%%%%%%%%%%%%%%%%%%%%%%%%%%%%%%%%%
This is known as the static-Gaussian-longitudinal-field Kubo-Toyabe (KT)
function. Figure \ref{SGLFKT} shows $\overline{P}_{z}(\omega_{\mathrm{L}%
},\Delta,t)$ for a variety of $\omega_{\mathrm{L}}$. Interestingly, despite
the fact that the external field is in the muon spin direction, wiggles are
seen in the polarization, and their frequency is given by $\omega_{\mathrm{L}%
}$. When $\omega_{\mathrm{L}}\gg\Delta$, the muon does not relax any more
because the field at the muon site is nearly parallel to the initial muon
spin direction. In this situation we say that the external field decoupled
the muon spin from the internal field. Finally, in the zero field case ($H_{%
\mathrm{L}}=0$) Eq.~\ref{GLF} reduces to \cite{HayanoPRB79}
\begin{equation}
\overline{P}_{z}(0,\Delta,t)=\frac{1}{3}+\frac{2}{3}(1-\Delta^{2}t^{2})\exp(-%
\frac{1}{2}\Delta^{2}t^{2}).  \label{GZF}
\end{equation}
This polarization function is known as the static-Gaussian-zero-field KT. At
early time it has a Gaussian-like behavior. It reaches a minimum on a time
scale set by $\Delta$ after which it recovers and saturates again to $1/3$.

\subsection{$\mu$SR: the dynamic case}

If we now add dynamics numerical methods must be applied. If the dynamic part of the local field at the muon site
$\delta \mathbf{B}$ fluctuates in time and magnitude so that
\begin{equation}
\left\langle \delta \mathbf{B}(t)\delta \mathbf{B}(0)\right\rangle =\frac{%
3\Delta ^{2}}{\gamma _{\mu }^{2}}\exp (-2\nu t),  \label{ClasicalFieldCorr}
\end{equation}%
where $\nu $ is a fluctuation rate, and under the strong collision
approximation, the muon polarization will obey the Volterra equation of the
second kind \cite{KerenJCMP04}. The polarization $\overline{P}_{z}(\nu
,\omega _{L}$,$\Delta ,t)$, which now also depends on the fluctuation rate $%
\nu $, obeys
\begin{eqnarray}
\overline{P}_{z}(\nu ,\omega _{L},\Delta ,t) &=&e^{-\nu t}\overline{P}%
_{z}(0,\omega _{L},\Delta ,t)+  \nonumber \\
&&\nu \int_{0}^{t}dt^{\prime }\overline{P}_{z}(\nu ,\omega _{L},\Delta
,t-t^{\prime })e^{-\nu t^{\prime }}\overline{P}_{z}(0,\omega _{L},\Delta
,t^{\prime })  \label{Voltera}
\end{eqnarray}%
where $\overline{P}_{z}(0,\omega _{L},\Delta ,t)$ is the static relaxation
function, namely, the polarization if the local field was frozen in time.
The factor $e^{-\nu t}$ is the probability to have no field changes up to
time $t$. The factor $e^{-\nu t^{\prime }}\nu dt^{\prime }$ is the
probability density to experience a field change only between $t^{\prime }$
and $t^{\prime }+dt^{\prime }$. The first term on the r.h.s is the
polarization at time $t$ due to muon that did not experienced any field
changes. The second term on the r.h.s is the contribution from those muon
that experience their first field change at time $t^{\prime }$. The factor $%
e^{-\nu t^{\prime }}\overline{P}_{z}(0,H,\Delta ,t^{\prime })\nu dt^{\prime
} $ is the amplitudes for the polarization function that evolves from time $%
t^{\prime }$ to $t$, which can involve more field changes recursively. This
equation can be solved numerically \cite{NumericalRecepes}.

%%%%%%%%%%%%%%%%%%%%FIG2%%%%%%%%%%%%%%%%%%%%%%%%%%%%%%%%%%%%%%%%%%%%
\begin{figure}[h!]
\vspace{6.5cm} \includegraphics{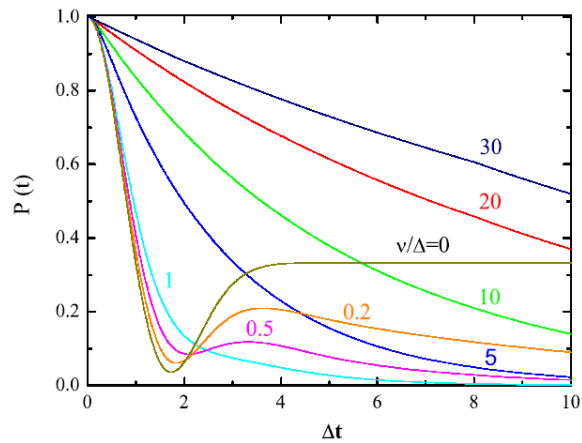}
\caption{Expected muon relaxation in a dynamic field with Gaussian
instantaneous distribution and no external field. Different values of
fluctuation rates are shown.}
\label{VoteraNu}
\end{figure}
%%%%%%%%%%%%%%%%%%%%%%%%%%%%%%%%%%%%%%%%%%%%%%%%%%%%%%%%%%%%%%%%%%%%%%%%%%%%%%

There are three ways of using Eq.~\ref{Voltera} to obtain dynamic
information. The first is in simple cases where $\overline{P}%
_{z}(0,\omega_{L},\Delta,t)$ is known analytically as was done by Brewer
\textit{et al.} \cite{BrewerHI86} for F-$\mu$-F bond. The second is when $%
\overline{P}_{z}(0,\omega_{L},\Delta,t)$ must be obtained numerically as in the cases of Gaussian
\cite{HayanoPRB79} or Lorenzian \cite{UemuraPRB85} field distribution with external longitudinal field. The third
way is to measure $\overline{P}_{z}(0,\omega_{L},\Delta,t)$ by cooling the system to low enough temperatures that
dynamic fluctuations are no longer present, and to use the measured $\overline{P}_{z}(0,\omega_{L},\Delta,t)$ as
an input to the Volterra equation \cite{KerenJCMP04}.

Taking the polarizations generated by static field distribution given by Eq.~%
\ref{GLF} with $\omega _{L}=0$, and using it as an input in the Volterra equation, gives the dynamic polarizations
shown in Fig. \ref{VoteraNu}. We see from this figure that when dynamic fluctuations are present the $1/3$
recovery is lost. As the fluctuation rate $\nu $ increases, the Gaussian-like relaxation at $t\rightarrow 0$ is
also lost, and when $\nu
>\Delta $ the relaxation becomes exponential.

\begin{figure}[h]
\vspace{7cm} \includegraphics{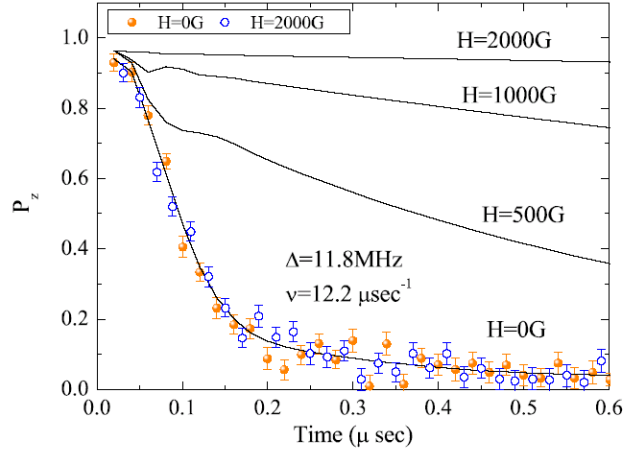}
\caption{Solid lines: Expected muon relaxation in a combination of internal
field fluctuations with Gaussian instantaneous distribution, and
longitudinal external field $H$. Symbols: Relaxation data from SrCr$_{9p}$Ga$%
_{12-9p}$O$_{19}$ from Ref.~\protect\cite{UemuraPRL94} in two different
values of $H$. The disagreement between data and model indicates unusual
behavior and is discussed in Sec.~2.5}
\label{SCGO}
\end{figure}

Finally, in Fig. \ref{SCGO} we present the most complicated relaxation function combining Gaussian field
distribution, fluctuations, and longitudinal field. This is known as the dynamic-Gaussian-longitudinal-field-KT
relaxation function. We have chosen a special value of the parameters $\Delta =11.8$~Mhz and $\nu =12.2$~MHz for
reasons that will become clear in Sec. 2.5, and show how the polarization is
modified as $H$ varies.%
%%%%%%%%%%%%%%%%%%%%FIG2%%%%%%%%%%%%%%%%%%%%%%%%%%%%%%%%%%%%%%%%%%%%
%%%%%%%%%%%%%%%%%%%%%%%%%%%%%%%%%%%%%%%%%%%%%%%%%%%%%%%%%%%%%%%%%%%%%%%%%%%%%%

It is interesting to mention that for the case $\nu \geq \Delta $ there is
an approximate expression for the dynamic-Gaussian-longitudinal-field-KT
relaxation function \cite{KerenPRB94} given by%
\begin{equation}
\overline{P}_{z}(t)=\exp (-\Gamma (t)t)  \label{KerenFunc}
\end{equation}%
where%
\[
\Gamma (t)t=\frac{2\Delta ^{2}\left\{ [\omega _{\mathrm{L}}^{2}+\nu ^{2}]\nu
t+[\omega _{\mathrm{L}}^{2}-\nu ^{2}][1-e^{-\nu t}\cos (\omega _{\mathrm{L}%
}t)]-2\nu \omega _{\mathrm{L}}e^{-\nu t}\sin (\omega _{\mathrm{L}}t)\right\}
}{(\omega _{\mathrm{L}}^{2}+\nu ^{2})^{2}}
\]%
In the long time limit ($\nu t\gg 1$) one finds that $\lim_{t\rightarrow
\infty }\overline{P}_{z}(t)=P_{0}\exp \left( -t/T_{1}\right) $ leading to
the standard expression for $T_{1}$
\begin{equation}
\frac{1}{T_{1}}=\frac{2\Delta ^{2}\nu }{(\omega _{\mathrm{L}}^{2}+\nu ^{2})}.
\label{InvT1}
\end{equation}%
This expression demonstrates that when the external field is small, namely, $%
\omega _{\mathrm{L}}\ll \nu $, $T_{1}$ will show no field dependence. But
when the field is large $\omega _{\mathrm{L}}\gg \nu $, $T_{1}$ will
increases with increasing field. Therefore, field dependent measurements can
be used to provide information on $\nu $, and distinguish between static and
dynamic cases.

Equation \ref{InvT1} could be related to Eq.~\ref{T1NMR} given for NMR $%
1/T_{1}$. In systems characterized by a spin-spin correlation function
decaying in time as $exp(-\nu t)$, neglecting the $q-$dependence, one can
write for the spin susceptibility
\[
\mathbf{\chi }_{\alpha \alpha }^{\prime \prime }(0,\omega )=\frac{(g\mu
_{B})^{2}}{\hbar V}\frac{\hbar \omega }{k_{B}T}N\left\langle S_{\alpha
}^{2}\right\rangle \frac{\nu }{\nu ^{2}+\omega ^{2}},
\]%
so that $\Delta ^{2}$ in Eq. 22 can be related to the amplitude of the spin
fluctuations
\[
\Delta ^{2}=\frac{\gamma ^{2}}{12}S(S+1)\sum_{\vec{q}}\left( |A_{\vec{q}%
}|^{2}\right) _{\perp }.
\]%
When the applied field is small or the fluctuations are fast Eq.~\ref{InvT1}
reduces to%
\begin{equation}
\frac{1}{T_{1}}=\frac{2\Delta ^{2}}{\nu }.  \label{ZFT1}
\end{equation}%
Typical values are $T_{1}\sim 0.1$~$\mu \sec ,$ $\Delta \sim 10$ MHz, and $%
\nu \sim 10$~$\mu \sec ^{-1}$.

\section{From zero to three-dimensional frustrated magnets}

\label{sec:2}

\subsection{Molecular Magnets}

In recent years major attention has been addressed to the study of molecular
crystals containing molecules formed by paramagnetic ions with significant
intramolecular exchange couplings and negligible intermolecular couplings,
so that each one of them can be considered as an independent nanomagnet \cite%
{Gatteschi}.
%%%%%%%%%%%%%%%%%%%%FIG2%%%%%%%%%%%%%%%%%%%%%%%%%%%%%%%%%%%%%%%%%%%%
\begin{figure}[h!]
\vspace{7cm} \includegraphics{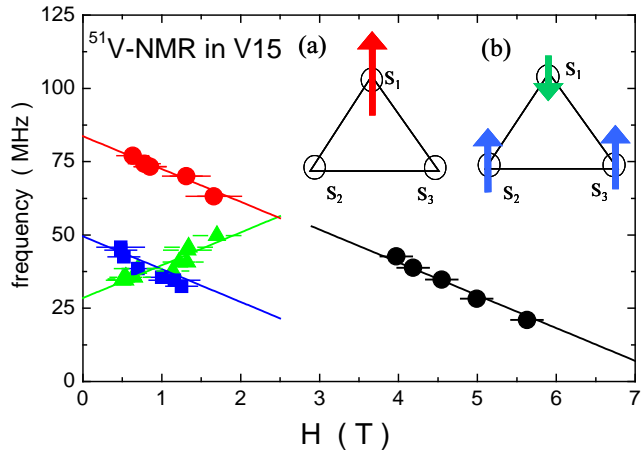}
\caption{Magnetic field dependence of the resonance frequency for different $%
^{51}$V sites in V$_{15}$, used to derive the spin polarization of each V$%
^{4+}$ \protect\cite{Furukawa}. In the inset the schematic view of the twofold degenerate ground-state is shown.}
\label{figfuru}
\end{figure}
%%%%%%%%%%%%%%%%%%%%%%%%%%%%%%%%%%%%%%%%%%%%%%%%%%%%%%%%%%%%%%%%%%%%%%%%%%%%%%

In some nanomagnets V$^{4+}$ $S=1/2$ ions form a triangular lattice. This is the case, for instance, of V$_{15}$,
where at low temperature (T) 12 V$^{4+}$ spins are coupled in singlets and the remaining 3 ions form a triangle.
From the study of $^{51}$V NMR shift below 100 mK Furukawa et al.\cite{Furukawa} have preliminarily estimated the
hyperfine coupling and then estimated the expectation values for V$^{4+}$ magnetic moments in zero-field. The
resonance frequency of the $i$-th nucleus is given by $\omega _{i}=\gamma (\vec{H}_{0}+\tilde{A}<\vec{S}_{i}>)$,
where the internal fields $\tilde{A}<\vec{S}_{i}>$ have a different orientation
and magnitude, depending on $<\vec{S}_{i}>$. In order to derive $<\vec{S}%
_{i}>$ the authors have studied the magnetic field dependence of $\omega
_{i} $, as shown in Fig. \ref{figfuru}. The expectation values were found to
be consistent with a doubly degenerate ground-state of the form
\begin{eqnarray}
\psi _{a} &=&\frac{1}{\sqrt{2}}(|\downarrow \downarrow \uparrow
>-|\downarrow \uparrow \downarrow >)  \nonumber \\
\psi _{a} &=&\frac{1}{\sqrt{6}}(|\uparrow \downarrow \downarrow
>-|\downarrow \uparrow \downarrow >-|\downarrow \downarrow \uparrow >)
\end{eqnarray}%
where $\uparrow $ or $\downarrow $ represents the orientation of each one of the three spins. $\mu $SR
longitudinal relaxation rate measurements in the same compound \cite{Proci} evidenced a nearly T-independent
relaxation rate at low temperature, which is possibly associated with the transitions among these two degenerate
ground-states. If one considers that the frequency of the spin fluctuations $\omega \gg \omega _{L}$, one can
derive the characteristic tunneling rate between the two degenerate states. We point out that a nearly
T-independent relaxation is observed also in frustrated magnets with a macroscopically degenerate ground-state
(see Sect. 2.6).

\subsection{Antiferromagnets on a square-lattice with competing
interactions: the $J_{1}-J_{2}$ model\label{J1J2}}

%%%%%%%%%%%%%%%%%%%%FIG2%%%%%%%%%%%%%%%%%%%%%%%%%%%%%%%%%%%%%%%%%%%%
\begin{figure}[h!]
\vspace{7.5cm} \includegraphics{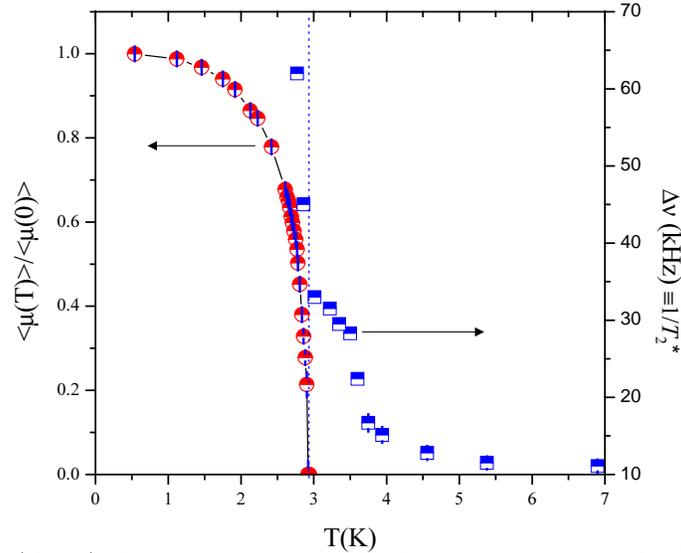} \caption{(circles)
Temperature dependence of the order parameter in the collinear phase of Li$_2$VOSiO$_4$, derived from zero-field
$\protect\mu$SR. (squares) Temperature dependence of $^7$Li NMR linewidth for $\vec H\parallel ab$ in
Li$_2$VOSiO$_4$. It is noticed that a broadening starts well above the transition temperature (vertical dotted
line) and is possibly due to a frustration driven lattice distortion.} \label{Mulisi}
\end{figure}
%%%%%%%%%%%%%%%%%%%%%%%%%%%%%%%%%%%%%%%%%%%%%%%%%%%%%%%%%%%%%%%%%%%%%%%%%%%%%%

V$^{4+}$ ions can also form other magnetic structures characterized by a
strong frustration of the magnetic moments. In fact, certain vanadates can
be considered as prototypes of frustrated magnets on a square lattice where
frustration arises from the competition between nearest neighbour ($J_1$)
and next-nearest neighbour ($J_2$) exchange couplings. The first NMR studies
have been carried out in Li$_2$VOSiO$_4$. $^7$Li NMR spectra were observed
to split in three different lines for $T< T_c\simeq 2.9$ K \cite%
{Melzi1,Melzi2}, one unshifted and the other two symmetrically shifted with
respect to the central one. This splitting of the NMR line was the first
evidence that this compound is characterized by a magnetic collinear
ground-state, as confirmed few years later by neutron scattering experiments
\cite{Bombardi}. A careful study of the order parameter has been carried out
by means of zero-field $\mu$SR measurements (Fig. \ref{Mulisi}), where the $%
\mu^+$ polarization is characterized by oscillations at a frequency directly
proportional to V$^{4+}$ magnetic moment \cite{MuLi} (Sect. 1). The
continuous increase of the order parameter for $T\rightarrow T_c$ was found
to be described by a critical exponent close to the one expected for 2D XY
universality class. Above $T_c$, when no internal field is present, the NMR
shift $K$ is expected to be proportional to the static uniform spin
susceptibility $\chi_s$ (see Eq. 5), with a slope given by the hyperfine
coupling. Remarkably, for $T_c< T< J_1+ J_2$ a change in the slope of $^7$Li
$K$ vs $\chi_s$ is noticed \cite{JPC}, suggesting a modification in the
average hyperfine coupling (see Sect. I). Moreover, the broadening of $^7$Li
NMR linewidth in the same temperature range can be explained only if the
distribution of the hyperfine couplings is also increasing in the same
temperature range. We remark that in this temperature range the average
shift decreases on cooling below 5 K while the linewidth increases, namely
the breakdown of Eq. 9 is noticed. A similar scenario is observed for $^{95}$%
Mo NMR in MoVO$_5$ \cite{MOVO}. These modifications in the hyperfine
coupling have been associated with a lattice distortion driven by the
spin-lattice coupling, which releases the degeneracy of the magnetic
ground-state.

%%%%%%%%%%%%%%%%%%%%FIG2%%%%%%%%%%%%%%%%%%%%%%%%%%%%%%%%%%%%%%%%%%%%
\begin{figure}[h!]
\vspace{6.4cm} \includegraphics{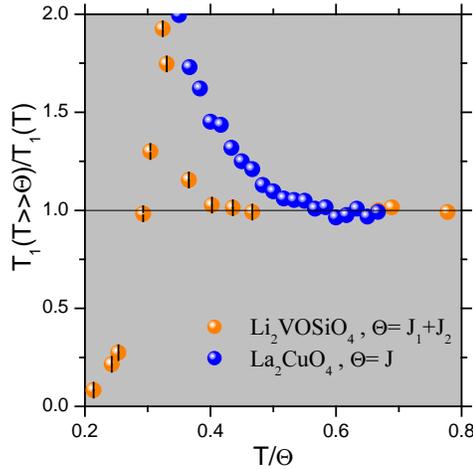}
\caption{Comparison of the temperature dependence of $^7$Li $1/T_1$ in Li$_2$%
VOSiO$_4$ with $^{63}$Cu $1/T_1$ in La$_2$CuO$_4$, where frustration is
negligible. The temperature is normalized to Curie-Weiss temperature $\Theta$%
, while $1/T_1$ is normalized to its value at $T\gg \Theta$. }
\label{t1lisivo}
\end{figure}
%%%%%%%%%%%%%%%%%%%%%%%%%%%%%%%%%%%%%%%%%%%%%%%%%%%%%%%%%%%%%%%%%%%%%%%%%%%%%%

The T-dependence of the in-plane correlation length $\xi $ of Li$_{2}$VOSiO$%
_{4}$ has been estimated from $^{7}$Li $1/T_{1}$ \cite{Papin}. In fact, by
resorting to scaling arguments one can rewrite the dynamical structure
factor in Eq. 12 in terms of powers of $\xi $. One finds
\begin{equation}
\frac{1}{T_{1}}\,\sim \,\xi ^{z}\,\sim \,\mathrm{exp}\,(2\,\pi z\,\rho
_{s}/T).  \label{scaling}
\end{equation}%
with $\rho _{s}$ the spin stiffness and $z$ the dynamical scaling exponent
which is estimated around $z=1$ on the basis of some physical considerations
\cite{Papin}. It turns out that for a two-dimensional antiferromagnet on a
square-lattice frustration yields a less pronounced increase of $\xi $ on
cooling, namely to a decrease in the spin stiffness.

The decay of the longitudinal muon polarization in Li$_2$VOSiO$_4$ and Li$_2$%
VOGeO$_4$ was observed to be well described by Eq. 18 and evidenced a
dynamics at frequencies well below the Heisenberg exchange frequency \cite%
{Papin}. This dynamic has been ascribed to the fluctuations within the
two-fold degenerate ground-state over a barrier \cite{Chandra}
\[
E(T)= \biggl({\frac{J_1^2S^2}{2J_2}}\biggr)\biggl[0.26\biggl({\frac{1}{S}}%
\biggr) + 0.318 \biggl({\frac{T}{J_2S^2}}\biggr)\biggr]\xi^2(T)\,\,\, .
\]

\subsection{Magnetic frustration on a triangular lattice}

\begin{figure}[h]
\vspace{9cm} \includegraphics{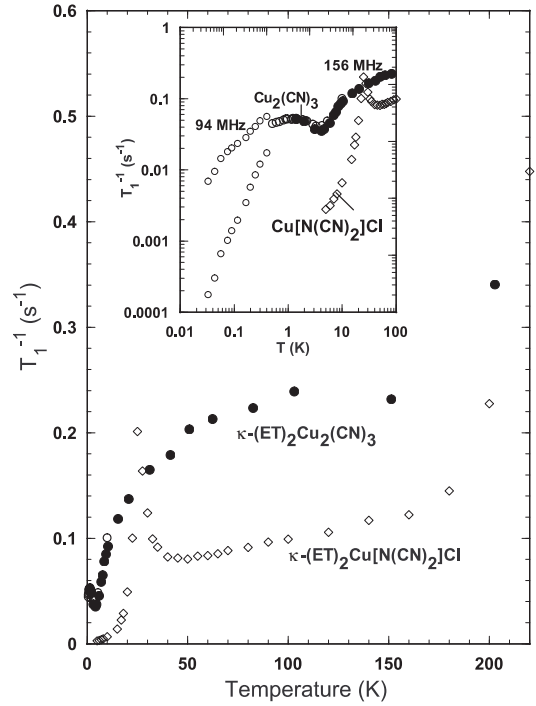}
\caption{$^{1}$H nuclear spin-lattice relaxation rate $1/T_{1}$ for a single
crystal (open circles) and a polycrystalline sample (closed circles) of $%
\protect\kappa $-(ET)$_{2}$Cu$_{2}$(CN)$_{3}$ and a single crystal of $%
\protect\kappa $-(ET)$_{2}$Cu[N(CN)$_{2}$]Cl (open diamonds) \protect\cite%
{Kanoda}. The Inset shows the low temperature part of the data in
logarithmic scales. }
\label{figkan1}
\end{figure}

The simplest two-dimensional lattice frustrated by the geometry of the
interactions is the triangular one. Several systems on a triangular lattice
have been investigated in the last decade, either insulating \cite{Triangle1}
or metallic ones \cite{Triangle2}. Some of these compounds display a rather
interesting phase diagram as a function of the magnetic field intensity, as
will be shown in a subsequent chapter of this book by M. Takigawa and F.
Mila. Also in $\kappa $-(ET)$_{2}$Cu$_{2}$(CN)$_{3}$ molecular crystals,
which are insulators at ambient pressure, the spins are arranged on a
triangular lattice. Recently, it has been observed (Fig.\ref{figkan1}) that $%
^{1}$H $1/T_{1}$ abruptly decreases at low temperature \cite{Kanoda}. This decrease can be associated with the
onset of a gap $\Delta $ between a collective singlet and triplet states, which yields $1/T_{1}\propto exp(-\Delta
/T)$ for $T\ll \Delta $, and can be considered the first evidence for a spin-liquid phase in a triangular
antiferromagnet. This could be possibly the experimental evidence of the long sought RVB state predicted by
Anderson long ago \cite{Anderson}. Interestingly enough the application of hydrostatic pressure was observed to
drive this system into a superconducting ground-state which can still be justified in terms of an RVB description.
%%%%%%%%%%%%%%%%%%%%FIG2%%%%%%%%%%%%%%%%%%%%%%%%%%%%%%%%%%%%%%%%%%%%
%%%%%%%%%%%%%%%%%%%%%%%%%%%%%%%%%%%%%%%%%%%%%%%%%%%%%%%%%%%%%%%%%%%%%%%%%%%%%%

Another spin system on a triangular lattice which has attracted much
attention in recent years is Na$_x$CoO$_2$, where a rich phase diagram
develops upon Na doping \cite{NaX}. An accurate study of the $^{23}$Na and $%
^{59}$Co NMR spectra in oriented powders evidenced a charge order for $%
x\simeq 0.7$ \cite{Alloul}. The signature of the charge order is the
presence of three distinct Na sites characterized by different quadrupole
couplings and magnetic shifts, which imply a well defined order of the Na$^+$
ions and of the Co charges in the CoO$_2$ planes. On the other hand, for $%
x=0.5$ the study of the temperature dependence of $^{59}$Co NMR spectra
reveals that the the electric field gradient at the Co site does not change
at the metal-insulator transition, indicating the absence of any charge
ordering \cite{Na05}. These NMR measurements have allowed to clarify the
nature of the ground-state of this system in the doping range $0.5\leq x\leq
0.75$ \cite{Gavi}. When full Na doping is achieved the system eventually
becomes non-magnetic and the NMR shift vanishes \cite{Lang}. Remarkably for $%
x=0$ the isostructural compound CoO$_2$ is found to be metallic, with a
crossover from a strongly correlated metal to a Fermi-liquid behaviour at
low temperature \cite{Marc}. In fact, around 4 K the temperature dependence
of $1/T_1T$ starts to flatten, as expected in a Fermi-liquid \cite{Slichter}.

%%%%%%%%%%%%%%%%%%%%FIG2%%%%%%%%%%%%%%%%%%%%%%%%%%%%%%%%%%%%%%%%%%%%
\begin{figure}[h]
\vspace{8.7cm} \includegraphics{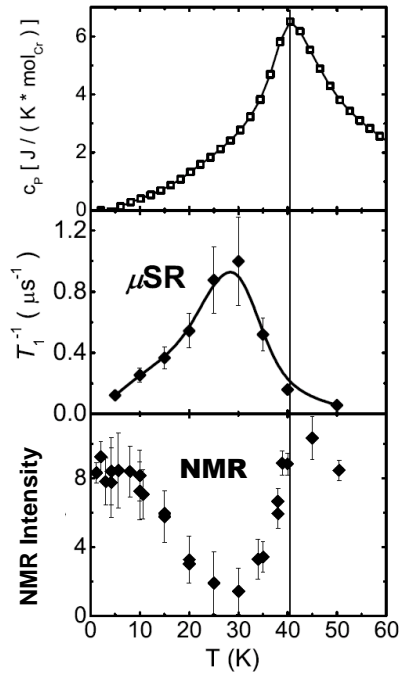} \caption{From the top:
specific heat, muon spin-lattice relaxation rate and $^{23}$Na NMR signal intensity in NaCrO$_2$ vs. temperature.
It is noticed that the peak in the specific heat and $1/T_1$ occur at different temperatures. The loss of
$^{23}$Na signal is possibly associated with an increase of the NMR relaxation rates.} \label{figolariu}
\end{figure}
%%%%%%%%%%%%%%%%%%%%%%%%%%%%%%%%%%%%%%%%%%%%%%%%%%%%%%%%%%%%%%%%%%%%%%%%%%%%%%
Recently, it has been pointed out that NaCrO$_2$ is an excellent realization of a $S=3/2$ triangular Heisenberg
antiferromagnet. Remarkably, in this compound while specific heat and magnetization measurements indicate the
onset of a transition around T$_c\simeq 40$ K, both muon spin rotation and NMR reveal a fluctuating crossover
regime extending well below T$_c$, with a peak in $1/T_1$ around 25 K \cite{Olariu2} (see Fig.\ref{figolariu}).
This apparent discrepancy might indicate the presence of vortex-antivortex excitations decoupling around 25 K.

Magnetic frustration can be associated not only with the geometry of the electron spin arrangement but also with
the one of indistinguishable nuclear spins, as it is the case for $^{3}$He. The triangular lattice topology can be
achieved by evaporating a single layer of $^{3}$He on a graphite substrate. Then one can conveniently use the
intensity of the NMR signal, which is proportional to the nuclear magnetization, to track the T dependence of the
nuclear spin susceptibility \cite{He3A}. $^{3}$He NMR measurements have been carried out down to tens of $\mu $K,
more than an order of magnitude below the exchange coupling. The low temperature increase in the nuclear spin
susceptibility evidenced the relevance of higher order multiple spin exchange interactions which, together with
the triangular geometry of the nuclear spins, makes the system strongly frustrated and possibly characterized by a
gapless spin-liquid ground-state \cite{He3B,He3A}.

%%%%%%%%%%%%%%%%%%%%%FIG2%%%%%%%%%%%%%%%%%%%%%%%%%%%%%%%%%%%%%%%%%%%%
\begin{figure}[h]
\vspace{5.5cm} \includegraphics{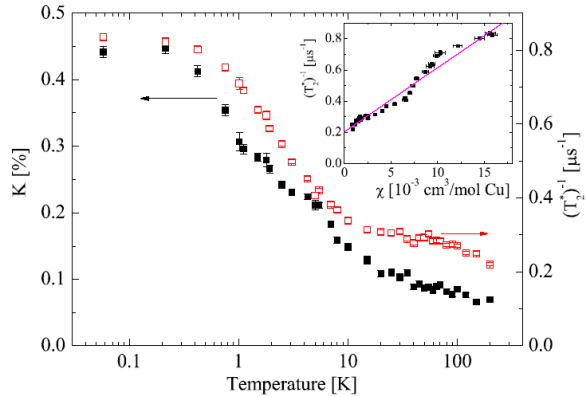}
\caption{The muon shift $K$ and transverse relaxation time $1/T_{2}^{\ast }$%
, versus temperature. Inset, a plot of $1/T_{2}^{\ast }$ versus the system's suspetibility.} \label{kagomehalf}
\end{figure}
%%%%%%%%%%%%%%%%%%%%%%%%%%%%%%%%%%%%%%%%%%%%%%%%%%%%%%%%%%%%%%%%%%%%%%%%%%%%%%

\subsection{$\protect\mu $SR and NMR in the spin-1/2 kagom\'{e} lattice ZnCu$%
_{3} $(OH)$_{6}$Cl$_{2}$}

A very interesting model compound for the study of magnetism on the kagom%
\'{e} lattice is the herbertsmithite ZnCu$_{3}$(OH)$_{6}$Cl$_{2}$. The
moments in this material originate from Cu$^{2+}$ which has a spin 1/2.
Therefore, it is ideal for the investigation of quantum ground states.
Unfortunately, different probes such as muon \cite{Ofer}, O, \cite{olariu}
Cu, and Cl \cite{imai} nuclear magnetic resonance suggest different behavior
of the shift below $\sim 50$~K and the origin of these variations are not
clear yet. Here we present only the $\mu $SR results. First we examine the
muon shift and $T_{2}^{\ast }$ in ZnCu$_{3}$(OH)$_{6}$Cl$_{2}$ independently
(Fig.~\ref{kagomehalf}), and also one as a function of the other (inset of
Fig.~\ref{kagomehalf}). The shift, and hence the susceptibility, increase
continuously upon cooling and saturate below 200~mK. This indicates that the
spin-1/2 kagom\'{e} does not freeze or form singlets. The ground state is
paramagnetic. In addition, the $K$ and $1/T_{2}^{\ast }$ follow each other
as the temperature is lowered, as shown in the inset, as expected from Eq. %
\ref{T2toK}. Although $K$ is not exactly a linear function of $1/T_{2}^{\ast
}$ there is no reason to suspect a modification in the hyperfine coupling
upon cooling, namely that lattice deformation is present in this case.%

\begin{figure}[h]
\vspace{5.5cm} \includegraphics{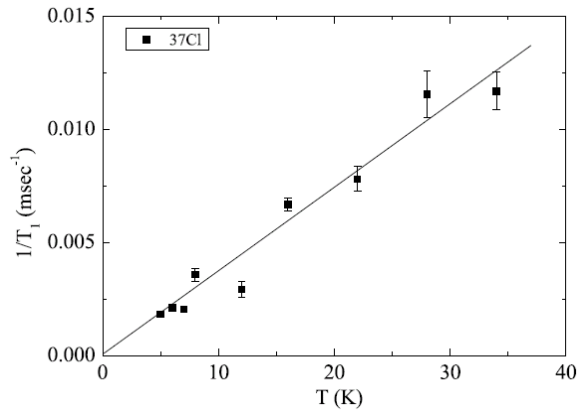}
\caption{$^{37}$Cl $1/T_{1}$ in the herbertsmithite ZnCu$_{3}$(OH)$_{6}$Cl$%
_{2}$ at low temperature.}
\label{Cl}
\end{figure}

Second we examine whether the ground state is separated by a gap from the
excited ones. If such a gap exists it will take a finite temperature to
generate excitations and achieve a non zero $\chi "_{\alpha \alpha }(\vec{q}%
,\omega _{0})$. Therefore, according to Eq.~\ref{T1NMR}, $1/(T_{1}T)$ should
extrapolate to zero. This has not been observed experimentally \cite%
{Herb,imai,olariu}. The $^{37}$Cl $1/T_{1}$ depicted in Fig.~\ref{Cl} is
proportional to $T$ meaning that $\chi "_{\alpha \alpha }(\vec{q},\omega
_{0})$ is finite in the $T\rightarrow 0$ limit. Therefore, there is no
evidence of a gap and the kagom\'{e} lattice seems to be an exotic magnet
with no broken continuous symmetry but gapless excitations.

%%%%%%%%%%%%%%%%%%%%%FIG2%%%%%%%%%%%%%%%%%%%%%%%%%%%%%%%%%%%%%%%%%%%%
%%%%%%%%%%%%%%%%%%%%%%%%%%%%%%%%%%%%%%%%%%%%%%%%%%%%%%%%%%%%%%%%%%%%%%%%%%%%%%

\subsection{The problem of the $\protect\mu ^{+}$ relaxation in some kagom%
\'{e} lattices}

\label{sec:2.5}

At the very beginning of the research in the field of frustrated magnets it
was noticed that the muon relaxation function is unusual \cite{UemuraPRL94}.
The symbols in Fig. \ref{SCGO} show the polarization at a temperature of
100~mK in the kagom\'{e} system SrCr$_{9p}$Ga$_{12-9p}$O$_{19}$ (SCGO) in
zero field and a longitudinal field of 2~kG \cite{UemuraPRL94}. First, no
oscillations are found, so the internal field is random with either static
or dynamic nature. Second, the relaxation at early time is Gaussian, with a
time scale of 0.1$~\mu $sec, so $\Delta $ must be on the order of 10~MHz.
Third, there is no recovery so there must be some dynamic as in Fig.~\ref%
{VoteraNu}. But it must be that $\nu \sim \Delta $. Had $\nu $ been larger
then $\Delta $, the initial relaxation would have been Lorenzian (See Eq.~%
\ref{KerenFunc} to \ref{InvT1} and Fig.~\ref{VoteraNu}). Had $\nu $ been
much slower, the polarization would have recovered at least partially. In
these circumstances a field of 2~kG which is equivalent to $\omega _{\mathrm{%
L}}=170$~MHz should have \textquotedblleft decoupled\textquotedblright\ the relaxation. This is not happening. The
solid lines in Fig. \ref{SCGO} represent the expected decoupling which is very different from the observed one. A
model has been proposed to explain this problem \cite{UemuraPRL94}, which received the name sporadic dynamic (SD)
\cite{BonoPRL}\cite{Yaouanc}.

\begin{figure}[h!]
\vspace{6cm} \includegraphics{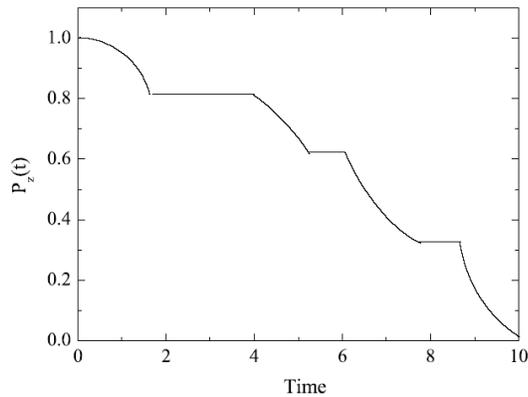}
\caption{Muon relaxation function in a case where the muons relax only
sporadically at certain intervals of time.}
\label{sd}
\end{figure}

In this model, the sample is not relaxing the muon spin all the time but
only for a fraction $f$ of the time, as demonstrated in Fig.~\ref{sd}. In
zero field it is clear that such a case will lead to sporadic dynamic
polarization $\overline{P}_{z}^{\mathrm{sd}}(\nu,0,\Delta,t)=\overline{P}%
_{z}(\nu ,0,\Delta,ft)$. However, even when the $H$ is applied, the
polarization changes only when the internal field relaxes the muon spin. In
other words, the flat parts in Fig.~\ref{sd} stay flat even when $H$ is on.
Therefore we expect $P_{z}^{\mathrm{sd}}(\nu,\omega_{\mathrm{L}},\Delta,t)=%
\overline{P}_{z}(\nu,\omega_{\mathrm{L}},\Delta,ft)$ for all values of $%
\omega_{\mathrm{L}}$. Since $\nu$, $\omega_{\mathrm{L}}$, and $\Delta$
always enter the relaxation function as a product with $t$ (for example see
Eq.~\ref{KerenFunc}) we must have $\overline{P}_{z}^{\mathrm{sd}%
}(\nu,\omega_{\mathrm{L}},\Delta,t)=\overline {P}_{z}(f\nu,f\omega_{\mathrm{L%
}},f\Delta,t)$.

%%%%%%%%%%%%%%%%%%%%%FIG2%%%%%%%%%%%%%%%%%%%%%%%%%%%%%%%%%%%%%%%%%%%%
%%%%%%%%%%%%%%%%%%%%%%%%%%%%%%%%%%%%%%%%%%%%%%%%%%%%%%%%%%%%%%%%%%%%%%%%%%%%%%

When analyzing the data we are actually estimating $f\Delta$ and $f\nu$ to
be 10~MHz and $\overline{P}_{z}^{\mathrm{sd}}=\overline{P}_{z}(10,f\omega _{%
\mathrm{L}},10,t)$. Therefore, the effect of the $H$ is reduced by factor $f$%
. In the fraction of the time the field is turned on, $\Delta$ and $\nu$ are
much higher than 10~MHz. This is the reason we do not see decoupling. This
model is very successful in explaining $\mu$SR data.
%%%%%%%%%%%%%%%%%%%%%FIG2%%%%%%%%%%%%%%%%%%%%%%%%%%%%%%%%%%%%%%%%%%%%
\begin{figure}[h!]
\vspace{7 cm} \includegraphics{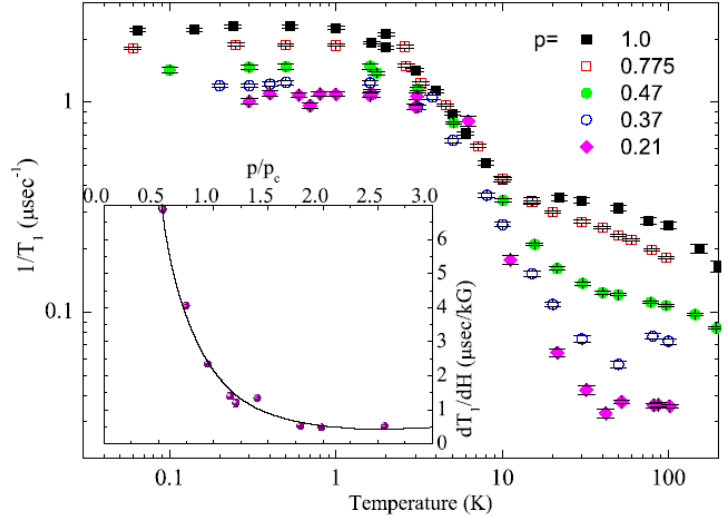} \caption{Temperature
dependence of the muon relaxation rate $1/T_{1}$ at H=50G. Inset, the rate at which $T_{1}$ varies with field as a
function of
magnetic ion concentration $p$ normalized by the percolation threshold $%
p_{c} $.} \label{tbtio}
\end{figure}
%%%%%%%%%%%%%%%%%%%%%%%%%%%%%%%%%%%%%%%%%%%%%%%%%%%%%%%%%%%%%%%%%%%%%%%%%%%%%%

The model could lead to two interpretations. The first one is that most of the time the field at the muon site is
zero due to singlet formation but every now and then a singlet breaks giving rise to a fluctuating field and
relaxation \cite{UemuraPRL94}. The problem with this interpretation is that the muon relaxation is
temperature-independent below 2-3K, similar to the case presented in Fig.~\ref{tbtio}. This usually happens when
the system is in its ground state. But in the ground state there cannot be time evolution, namely, the field
cannot turn on and off and the muon spin cannot relax. If the system is not in the ground state at $T\simeq 2$ K,
then a lower energy scale, below about 100 mK, must exist which separates the ground from the first excited
states. Accordingly, the constant relaxation observed for $100 \mathrm{mK}\leq T\leq 2$ K, would arise from the
quantum fluctuations within a manifold of weakly coupled nearly degenerate states. Another possibility is that the
muon is hopping, although not evenly, between two sites with different relaxation rates. This, however, is also
unusual at such low temperatures. A full description of the muon spin behavior in SCGO and similar systems is
still lacking.

\subsection{Persistent dynamics and lattice distortions in the pyrochlore
lattice}

If there is one common message deducible from the study of frustrated magnets by $\mu $SR it is that these systems
maintain a fluctuating part of the moment even when the temperature is lowered much below the coupling energy
scale. This is manifested in the saturation of $1/T_{1}$ upon cooling. In zero field $1/T_{1}$ is given by
Eq.~\ref{ZFT1}. Therefore, for $1/T_{1}$ to remain finite, at least part of the moment must continue to fluctuate
and $\Delta $ cannot be zero. While both $T_{1}$ and $\Delta $ can vary from system to system, the typical scale
for $\nu $ is $\mu \sec ^{-1}$.

In Fig.~\ref{tbtio} we present what might be the canonical example of
persisting fluctuations to the lowest temperatures, Tb$_{2}$Ti$_{2}$O$_{7}$
\cite{DunsigerPRB96}. The full black symbols denote $1/T_{1}$ over a wide
temperature range. The increase in $1/T_{1}$ upon cooling indicates that $%
\nu $ decreases, namely, the spin fluctuations slow down. However, in
standard magnets, at some temperature long range order or spin freezing sets
in, and the amplitude of the fluctuations $\delta B$ (see Eq.~\ref%
{ClasicalFieldCorr}) also decrease upon cooling. This is manifested in a
decrease in $\Delta ^{2}$ upon cooling. The net result is a peak in $1/T_{1}$
at, or close to, the critical temperature. The $1/T_{1}$ peak is missing in
Tb$_{2}$Ti$_{2}$O$_{7}$ suggesting that no static magnetic moment develops
in this system. Similar results where obtained in other pyrochlore lattices
such as Tb$_{2}$Sn$_{2}$O$_{7}$ \cite{Dalamas} and Gd$_{2}$Ti$_{2}$O$_{7}$
\cite{Yaouanc}.

This conclusion leads to yet another open question in this area of research, namely, which type of excitation will
dominate at low temperatures: spin wave, spinless, or spinons? This can be examined by measuring $1/T_{1}$ as a
function of magnetic ion concentration $p$ above and below the percolation threshold $p_{c}$. A strong dependence
of $1/T_{1}$ on $p$ close to $p_{c}$ would suggest that the fluctuations emerge from a collective phenomenon. In
contrast, if $1/T_{1}$ varies smoothly across $p_{c}$ it would suggest that the excitations are local in nature
and impartial to the coverage of the
lattice. In Fig.~\ref{tbtio} $1/T_{1}$ is presented for (Tb$_{p}$Y$_{1-p}$)$%
_{2}$Ti$_{2}$O$_{7}$ samples in which the Tb magnetic ion is replaced by
non-magnetic Y \cite{KerenPRL04}. Clearly the fluctuations have similar
behavior both above and below the percolation threshold which is at $%
p_{c}=0.39$ \cite{HenelyCJP01}.

Moreover, the muon $T_{1}$ was found to be a linear function of $H$ \cite%
{KerenPRL04}. This is a very different behavior from Eq.~\ref{InvT1} and suggests that the field correlation does
not decay exponentially but rather with a power law. In this case, the dynamic properties of the system could
be collected into $dT_{1}/dH$. In the inset of Fig.~\ref{tbtio} we show $%
dT_{1}/dH$ as a function of $p$. This quantity shows no anomaly at $p_{c}$,
suggesting that $T_{1}$ is controlled be local excitations.

%%%%%%%%%%%%%%%%%%%%%FIG2%%%%%%%%%%%%%%%%%%%%%%%%%%%%%%%%%%%%%%%%%%%%%%%%%%%%
\begin{figure}[h!]
\vspace{6cm} \includegraphics{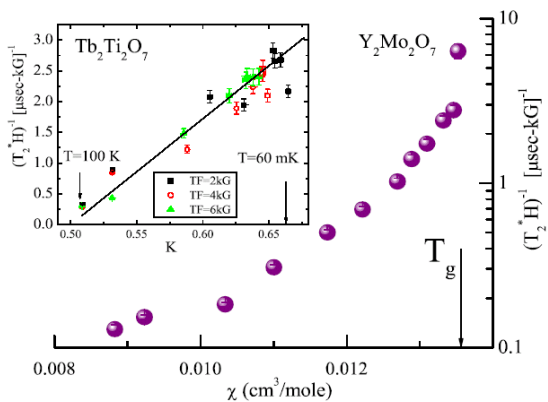}
\caption{$1/T_{2}^{\ast}$ normalized by the field \ for Y$_{2}$Mo$_{2}$O$%
_{7} $ versus susceptibility on a semi-log scale, with the temperature an an
implicit parameter. Inset, $1/T_{2}^{\ast}$ normalized by the field \ for Tb$%
_{2}$Ti$_{2}$O$_{7}$ on a linear scale.}
\label{ymoo}
\end{figure}
%%%%%%%%%%%%%%%%%%%%%%%%%%%%%%%%%%%%%%%%%%%%%%%%%%%%%%%%%%%%%%%%%%%%%%%%%%%%%%

Finally, as in the case of Li$_{2}$VOSiO$_{4}$ (see Sec.~\ref{J1J2}),
magnetoelastic coupling is a very important interaction \cite%
{Distortion,KerenPRL01} in the pyrochlore system as well. It could lead to a
lattice deformation in order to relieve the frustration. In fact, regardless
of how small this interaction is, it will cause some lattice deformation. In
other words, there is no such thing as a perfect Heisenberg pyrochlore
lattice since it must distort. Of course, for very small magnetoelastic
coupling the distortion could be undetectably small. Nevertheless, the
investigation of the ground state in the presence of magnetoelastic coupling
is a growing theoretical sub-field accompanied by experimental search for
this effect.

As explained above, a comparison between the muon relaxation rate and the
muon shift or macroscopic susceptibility can provide a hint about lattice
deformations. For example, in the pyrochlore Y$_{2}$Mo$_{2}$O$_{7}$, $%
1/T_{2}^{\ast}$ depends exponentially on the susceptibility. This is
demonstrated in Fig.~\ref{ymoo} on a semi-log scale. The fact that the
lattice parameters in Y$_{2}$Mo$_{2}$O$_{7}$ vary upon cooling received
confirmation also from NMR \cite{KerenPRL01} and x-ray \cite{OferInPrep}
experiments. In contrast, the lattice of Tb$_{2}$Ti$_{2}$O$_{7}$ does not
distort \cite{HanPRB04}, and, indeed, $1/T_{2}^{\ast}$ depends linearly on
the susceptibility as shown in the inset of Fig.~\ref{ymoo}.

\section*{Acknowledgements}

The authors would like to acknowledge Y. Furukawa and K. Kanoda for
contributing to this manuscript with the figures reporting their results. A.
Keren acknowledges support by the Israel U.S.A. binational science
foundation.


\begin{thebibliography}{99}
\bibitem{Slichter} C. P. Slichter in \textit{Principles of Magnetic Resonance%
} (Springer, Berlin, 1990) 3rd Ed.; A. Abragam in \textit{Principles of
Nuclear Magnetism} (Oxford University Press, New York, 1961).

\bibitem{Schenck} Muon science, edited by S. L. Lee, S. H. Kilcoyne and R.
Cywinski (IOP Publishing, Bristol and Philadelphia, 1999); {\textrm{A.}~%
\textrm{Schenck}}, \emph{Muon Spin Rotation: Principles and Applications in
Solid State Physics} ({Adam Hilger}, {Bristol}, {1985}); {\textrm{S.}~%
\textrm{Blundell}}, {Contemporary Physics 40} \textbf{40}, {175} ({1999}).

\bibitem{Slic2} C.H. Pennington and C.P. Slichter, Phys. Rev. Lett. 66, 381
(1991).

\bibitem{OferPRB06} R.Ofer, S. Levy, A. Kanigel, and A. Keren, Phys. Rev. B
\textbf{73}, 012503, (2006).

\bibitem{Alloul} I. R. Mukhamedshin, H. Alloul, G. Collin, and N. Blanchard,
Phys. Rev. Lett. 93, 167601 (2004)

\bibitem{Limot} L. Limot, P. Mendels, G. Collin, C. Mondelli, B. Ouladdiaf, H. Mutka, N. Blanchard, and M. Mekata,
Phys. Rev. B \textbf{65}, 144447 (2002)

\bibitem{Tedoldi} see for instance H. Alloul, J. Bobroff, M. Gabay, and P. J. Hirschfeld,
Rev. Mod. Phys. \textbf{81}, 45 (2009) and references therein


\bibitem{HayanoPRB79} R. S. Hayano, Y. J. Uemura, J. Imazato, N. Nishida, T.
Yamazaki, and R. Kubo, Phys. Rev. B \textbf{20}, 850 (1979).

\bibitem{KerenPRB93} A. Keren, L. P. Le, G. M. Luke, B. J. Sternlied, W. D.
Wu, Y. J. Uemura, S. Tajima, S. Uchida, Phys. Rev. B \textbf{48}, 12926
(1993).

\bibitem{NumericalRecepes} W. H. Press, B. P. Flannery, A. A. Teukolsky, and
W. T. Vetterling, \textit{Numerical Recipes} (Cambridge University Press,
Cambridge, 1989).

\bibitem{BrewerHI86} J. H.\ Brewer, D.R. Harshman, R. Keitel, S. R.
Kreitzman, G. M. Luke, D. R. Noakes, and R. \ E. Turner, Hyperfine
Interactins \textbf{32}, 677 (1986).

\bibitem{UemuraPRB85} Y. J. Uemura T. Yamazaki D. R. Harshman M. Senab, and
E. J. Ansaldo, Phys. Rev. B \textbf{31} (1985).

\bibitem{KerenJCMP04} A. Keren, J. Phys.: Condens. Matter \textbf{16}, 1
(2004).

\bibitem{KerenPRB94} A. Keren, Phys. Rev. B \textbf{50}, 10039 (1994).

\bibitem{Gatteschi} D. Gatteschi, A. Caneschi, L. Pardi, and R. Sessoli,
Science 265, 54 (1994)

\bibitem{Furukawa} Y. Furukawa, Y. Nishisaka, K. Kumagai, P. K\"ogerler, and
F. Borsa, Phys. Rev. B 75, 220402 (2007)

\bibitem{Proci} D. Procissi, A. Lascialfari, E. Micotti, M. Bertassi, P.
Carretta, Y. Furukawa, and P. K\"ogerler, Phys. Rev. B 73, 184417 (2006)

\bibitem{Melzi1} R. Melzi, P. Carretta, A. Lascialfari, M. Mambrini, M.
Troyer, P. Millet and F. Mila, Phys. Rev. Lett. 85, 1318 (2000)

\bibitem{Melzi2} R. Melzi, S. Aldrovandi, F. Tedoldi, P. Carretta, P. Millet
and F.Mila, Phys. Rev. B 64, 024409 (2001)

\bibitem{Bombardi} A. Bombardi, F. de Bergevin, S. Di Matteo, L. Paolasini,
P.Carretta, J. Rodriguez-Carvajal, P. Millet and R. Caciuffo, Phys.Rev.
Lett. 93, 027202 (2004)

\bibitem{MuLi} P. Carretta, R. Melzi, N. Papinutto and P. Millet, Phys. Rev.
Lett. 88, 047601 (2002)

\bibitem{JPC} P. Carretta, N. Papinutto, R. Melzi, P. Millet, S. Gouthier,
P. Mendels and P. Wzietek, J. Phys.: Condens. Matter 16, S849-S856 (2004)

\bibitem{MOVO} P.~Carretta, N.~Papinutto, C.~B.~Azzoni, M.~C.~Mozzati,
E.~Pavarini, S. Gonthier, and P. Millet, Phys. Rev. B \textbf{66}, 094420
(2002)

\bibitem{Papin} N. Papinutto, P. Carretta, S. Gonthier and P. Millet, Phys.
Rev. B 71, 174425 (2005)

\bibitem{Chandra} P. Chandra, P. Coleman and A. I. Larkin, Phys. Rev. Lett.
64, \textbf{88} (1990)

\bibitem{Triangle1} L. K. Alexander, N. B\"uttgen, R. Nath, A. V. Mahajan
and A. Loidl, Phys. Rev. B 76, 064429 (2007); A. Olariu, P. Mendels, F.
Bert, B. G. Ueland, P. Schiffer, R. F. Berger, and R. J. Cava, Phys. Rev.
Lett. 97, 167203 (2006)

\bibitem{Triangle2} Y. Kurosaki, Y. Shimizu, K. Miyagawa, K. Kanoda, and G.
Saito, Phys. Rev. Lett. 95, 177001 (2005); A. Kawamoto, Y. Honma, and K.
Kumagai, Phys. Rev. B 70, 060510 (2004)

\bibitem{Kanoda} Y. Shimizu, K. Miyagawa, K. Kanoda, M. Maesato, and G.
Saito, Phys. Rev. Lett. 91, 107001 (2003)

\bibitem{Anderson} P.W. Anderson, Mater. Res. Bull. 8, 153 (1973)

\bibitem{NaX} M. L. Foo, Y. Wang, S. Watauchi, H. W. Zandbergen, T. He, R.
J. Cava, and N. P. Ong, Phys. Rev. Lett. 92, 247001 (2004)



\bibitem{Na05} J. Bobroff, G. Lang, H. Alloul, N. Blanchard, and G. Collin,
Phys. Rev. Lett. 96, 107201 (2006)

\bibitem{Gavi} J. L. Gavilano, B. Pedrini, K. Magishi, J. Hinderer, M.
Weller, H. R. Ott, S. M. Kazakov, and J. Karpinski, Phys. Rev. B 74, 064410
(2006)

\bibitem{Lang} G. Lang, J. Bobroff, H. Alloul, P. Mendels, N. Blanchard, and
G. Collin, Phys. Rev. B 72, 094404 (2005)

\bibitem{Marc} C. de Vaulx, M.-H. Julien, C. Berthier, S. H\"ubert, V.
Pralong, and A. Maignan, Phys. Rev. Lett. 98, 246402 (2007)

\bibitem{Olariu2} A. Olariu, P. Mendels, F. Bert, B. G. Ueland, P. Schiffer, R. F. Berger, and R. J. Cava,
Phys. Rev. Lett. \textbf{97}, 167203 (2006)

\bibitem{He3A} H. Ikegami, R. Masutomi, K. Obara and H. Ishimoto, Phys. Rev.
Lett. 85, 5146 (2000)

\bibitem{He3B} M. Roger, C. B\"{a}uerle, Yu. M. Bunkov, A.-S. Chen and H.
Godfrin, Phys. Rev. Lett. 80, 1308 (1998); H. Ishimoto, R. Masutomi, H.
Ikegami, Y. Karaki and A. Yamaguchi, J. Phys. Chem. Solids 66, 1417 (2005).

\bibitem{Ofer} O. Ofer and A. Keren, accepted to PRB.

\bibitem{olariu} A. Olariu, P. Mendels, F. Bert, F. Duc, J. C. Trombe, M. A.
de Vries, and A. Harrison, Phys. Rev. Lett. \textbf{100} 087202 (2008).

\bibitem{imai} T. Imai, E. A. Nytko, B. M. Bartlett, M. P. Shores and D. G.
Nocera, Phys. Rev. Lett. \textbf{100} 077203, (2008).

\bibitem{Herb} O. Ofer, A. Keren, E. A. Nytko, and M. P. Shores,
cond-matt/0610540.

\bibitem{UemuraPRL94} Y. J. Uemura \textit{et al.}, Phys. Rev. Lett. \textbf{%
73}, 3306 (1994).

\bibitem{BonoPRL} D. Bono, \textit{et al.} Phys. Rev. Lett. \textbf{93},
187201 (2004).

\bibitem{Yaouanc} A. Yaouanc, Physica B \textbf{374-375}, 145-147 (2006).;

\bibitem{HenelyCJP01} C. L. Henley, Can. J. Phys. \textbf{79}, 1307 (2001).

\bibitem{DunsigerPRB96} S. R. Dunsiger et al., Phys. Rev. B \textbf{54},
9019 (1996).

\bibitem{Dalamas} P. Dalmas de R\'{e}otier, A. Yaouanc, L. Keller, A.
Cervellino, B. Roessli, C. Baines, A. Forget, C. Vaju, P. C. M. Gubbens, A.
Amato, and P. J. C. King, Phys. Rev. Lett. \textbf{96}, 127202 (2006).

\bibitem{Alan} A. Yaouanc, P. Dalmas de R\'{e}otier, V. Glazkov, C. Marin,
P. Bonville, J. A. Hodges, P. C. M. Gubbens, S. Sakarya, and C. Baines,
Phys. Rev. Lett. \textbf{95}, 047203 (2005).

\bibitem{KerenPRL04} A. Keren,. S. Gardner, G. Ehlers, A. Fukaya, E. Segal,
and Y. J. Uemura, Phys. Rev. Lett. \textbf{92}, 107204 (2004).

\bibitem{Distortion} K. Terao, J. Phys. Soc. Japan \textbf{65}, 1413 (1996);
Y. Yamashita and K. Ueda, Phys. Rev. Lett. \textbf{85}, 4960 (2000);
S.-H.~Lee, C.~Broholm, T.~H.~Kim, W.~Ratcliff II, and S-W.~Cheong, Phys.
Rev. Lett. \textbf{84}, 3718 (2000); O. Tchernyshyov, R. Moessner, and S. L.
Sondhi, Phys. Rev. Lett. \textbf{88}, 067203 (2002); J. Richter, O. Derzhko,
and J Schulenburg, Phys. Rev. Lett. \textbf{93}, 107206 (2004); D. L.
Bergman, R. Shindou, G. A. Fiete, and L. Balents, Phys. Rev. B \textbf{74},
134409 (2006), K. Penc, N. Shannon, and Hiroyuki Shiba, Phys. Rev. Lett.
\textbf{93}, 197203 (2004) ; F. Wang and A. Vishwanath, cond-mat/0709.3546.

\bibitem{KerenPRL01} A. Keren and J. S. Gardner, Phys. Rev. Lett. \textbf{87}%
, 177201 (2001).

\bibitem{OferInPrep} O. Ofer \textit{et al.} in preparation.

\bibitem{HanPRB04} S.-W. Han, J. S. Gardner, and C. H. Booth, Phys. Rev. B
\textbf{69}, 024416 \ (2004).
\end{thebibliography}
\end{document}